\documentclass[20pt]{article}

\usepackage{amsmath,amsthm,amssymb,
	amscd,float, setspace,wrapfig,
	amsfonts, url, subfigure}
\usepackage{bm, graphicx, color}
\newcommand{\bu}{{\mathbf u}}

 \def\bn{\mathbf n}

\def\be{{\boldsymbol e}}

\newcommand{\cW}{\mathcal{W}}
\newcommand{\mm}{\mathbf m}
\newcommand{\br}{\mathbf r}

\newcommand\tr{\operatorname{tr}}

\newcommand{\Energy}{\mathcal{E}}

\newcommand{\bp}{\mathbf  p}

\def\tr{\mathop{\rm tr}\nolimits}

\newcommand{\Om}{{\Omega}}

\newcommand{\bx}{{\mathbf x}}

\newcommand{\Ccchamber}{C_{\textrm{\begin{tiny}C\end{tiny}}}^{\textrm{\begin{
tiny}II\end{tiny}}}}

\newcommand{\Chcominus}{C_{{\textrm{\begin{tiny}H\end{tiny}}}{\textrm{\begin{
tiny}CO\end{tiny}}}_3^{\textrm{\begin{tiny}-\end{tiny}}}}}

\newcommand{\Jhcominus}{J_{{\textrm{\begin{tiny}H\end{tiny}}}_2{\textrm{\begin{
tiny}CO\end{tiny}}}_3^{-}}^{\textrm{\begin{tiny}II\end{tiny}}}}

\newcommand{\bnu}{{\boldsymbol\nu}}

\newcommand{\Curl}{\mathrm{curl}\,}       
\newcommand{\Div}{\mathrm{div}\,}         

\newcommand{\vn}{\mathbf{n}}

\newcommand{\vnu}{\bm{\nu}}

\newcommand{\vV}{\mathbf{V}}

\newcommand{\Gm}{\Gamma}

\newcommand{\iO}{\int_{\Omega}}

\newcommand{\iG}{\int_{\Gamma}}

\newcommand{\Eerk}{E_{\mathrm{erk}}}
\newcommand{\Echr}{E_{\mathrm{chr}}}
\newcommand{\Eelas}{E_{\mathrm{elas}}}

\newcommand{\Bulkfunc}{w}
\newcommand{\Bulkcoef}{\gamma_{0}}
\newcommand{\Bulkhatcoef}{\hat{\gamma}_{0}}

\newcommand{\Eanch}{E_{\mathrm{a}}}
\newcommand{\Ewas}{E_{\mathrm{a},s}}
\newcommand{\Ewan}{E_{\mathrm{a},\vn}}
\newcommand{\aperp}{a_{\perp}}
\newcommand{\apar}{a_{\parallel}}

\newcommand{\aori}{a_{\mathrm{ori}}}
\newcommand{\anchcoef}{a_{0}}
\newcommand{\anchhatcoef}{\hat{a}_{0}}

\newcommand{\Sas}{s_{\mathrm{a}}}

\newcommand{\bbm}{\mathbf{m}}
\newcommand{\bbe}{\mathbf{e}}

\begin{document}
\setcounter{page}{1}
\setcounter{section}{0}

\author {Shawn Walker\\Department of Mathematics,
		Louisiana State University,\\
		Baton Rouge, LA 70803
		\\ \\Javier Arsuaga\\Department of Cellular and Molecular Biology and Department of Mathematics,\\
		 University of California Davis, Davis, CA 95616\\ \\
		M.Carme Calderer\\ School of Mathematics, University of Minnesota,\\ Minneapolis, MN 55455\\ \\
Lindsey Hiltner\\School of Mathematics, University of Minnesota,\\ Minneapolis, MN 55455\\
\\ Mariel Vazquez\\Department of Microbiology and Molecular Genetics,\\ University of California Davis, Davis, CA 95616
}

\title{Fine Structure of Viral ds DNA Encapsidation}
\bibliographystyle{siam}
\date{\today}
\maketitle
\thanks{The authors are very grateful to this support granted to the project  by the National Science Foundation through grants NSF-DMS-1817156(Arsuaga and Vazquez), NSF-DMS-1816740 (Calderer) and NSF-DMS-1555222-CAREER (Walker)}
\begin{abstract}
	{\it In vivo} configurations of dsDNA of bacteriophage viruses in a capsid are known to form  hexagonal  chromonic liquid crystal phases. This article studies the liquid crystal ordering  of viral dsDNA in an icosahedral capsid, combining the chromonic model with that of liquid crystals with variable degree of orientation. The scalar order parameter of the latter allows us to distinguish  regions of the capsid with well-ordered DNA from the disordered central core.  We employ a state-of-the-art numerical algorithm based on the finite element method to find equilibrium states of the encapsidated DNA and calculate the corresponding pressure. With a  data-oriented  parameter selection strategy, the method yields phase spaces of the pressure and the radius of the disordered core, in terms of relevant dimensionless parameters, rendering the proposed algorithm into a preliminary bacteriophage designing tool. The presence of the order parameter also has the unique role of allowing for 	non-smooth capsid domains as well as accounting for knot locations of the DNA.
\end{abstract}

	
	\maketitle

	
	\section{Introduction}
	The discovery of liquid crystal phases formed by DNA in free solution as well as inside viral capsids led to the {\it chromonic} denomination of a class of lyotropic liquid crystals.  Different from liquid crystals consisting of elongated, rod-like molecules found, for instance, in display devices, chromonic liquid crystals consist of disk, plank-like molecules that form liquid crystal phases with varying concentration. The molecules stack to form cylindrical aggregates, and at even higher concentration, the cylinders cluster so their center axes align  in a hexagonal lattice, forming what is known as the hexagonal columnar liquid crystal phase.  Experimental and theoretical studies acquired over the last 30 years  \cite{leforestier2009structure, leforestier2010bacteriophage,lepault1987organization,reith2012effective,
		rill1986liquid,
		strzelecka1988multiple, park2008self} have shown that encapsidated DNA molecules form a columnar  hexagonal liquid crystal phase. In particular, liquid crystalline phases in bacteriophages were first proposed in \cite{kellenberger1986considerations}, with an explicit reference to 
	hexagonal packing made in \cite{livolant1991ordered}
	and since then, consistent data have been accumulating \cite{leforestier1993supramolecular,leforestier2008bacteriophage,marenduzzo2009dna,reith2012effective}.

	This article studies the liquid crystal ordering  of viral dsDNA in a capsid with axial symmetry. We use advanced liquid crystal models and their state-of-the-art numerical implementation to determine the equilibrium structure of the DNA in the capsid and the forces that it sustains. The model is also aimed to be used as a tool for the design and prediction of viral encapsidated DNA in response to the renewed interest in  bacteriophages in medicine and biotechnology. This goal  guides our parameter selection strategy which is also a main component of the work. The  proposed method delivers a parameter phase space for the pressure and the the size of the inner disordered core.  
	
	Icosahedral bacteriophages consist of a protein capsid  whose assembly is followed by the packing, by means of a molecular motor,  of a single naked dsDNA molecule  \cite{smith2001bacteriophage}.  The DNA molecule inside the viral capsid is found under extreme concentration,  between $200$ and $800$ mg/ml \cite{kellenberger1986considerations},  and pressure estimated in the  range between 40 and 60 atmospheres \cite{Evilevitch2003,Jeembaeva2008}. Three factors contribute to the excess pressure found inside the viral capsid:  the decrease in entropy associated with the confinement imposed by the capsid, the high resistance of the DNA molecule to bending beyond its persistence length and the self-repulsion of the DNA molecule \cite{RiemerBloomfield1978}. At the time of infection the DNA is released by a mechanism that suggests a phase transition, possibly into a 'liquid-like' state \cite{leforestier2010bacteriophage, liu2014solid, sae2014solid}. Both processes, packing and releasing of the genome depend greatly on how the DNA molecule folds within the viral capsid; however our understanding of this folding remains very limited.

	We develop an effective mechanical model of DNA encapsidation based on the theory of hexagonal chromonic liquid crystals, combined with that of liquid crystals with variable degree of orientation. 
	We characterize the hexagonal columnar phase in terms of a triple of orthonormal vectors, $\bn$, $\bbm$ and $\bp$,  the first giving the local direction of the columnar axis, with $\bbm$ and $\bp$ denoting lattice vectors of the orthogonal cross sections. The unit vector $\bn$ also represents the tangent vector to the DNA center curve at a point, with $\bbm$ and $\bp$ representing the directions  along which neighboring strands are found in the encapsidated arrangement. 
	Moreover, we include the scalar order parameter $s\in (0, 1)$ that describes the degree of orientation of filaments in the capsid \cite{ericksen1991degree}. The value  $s=1$ corresponds to (ideal) perfect alignment and $s=0$ to  the isotropic state, with no alignment ordering, so it naturally lends itself to quantitatively distinguishing between the region of the capsid with ordered DNA from the disordered central core. We point out that $\bn, \bbm $ and $\bp $ are undefined at points where $s=0$. The order parameter in the model also yields a natural realization of the phase transition from the ordered to the liquid-like state. Furthermore,  the model allows for the capsid domain to be of icosahedral shape, since $s=0$ at the edges where the vector fields are not defined.  
		Ordering of the DNA molecule at the boundary as promoted by the capsid \cite{Earnshaw1980,Effantin2006,lepault1987organization,Cerritelli1997} motivates our selection of the boundary conditions for the fields of the problem.

	In addition to the genome length $L$, and the volume $V$ of the capsid, data at our disposal include cry-EM images of the bacteriophage yielding the size and shape of the inner disordered region,  the DNA density graphs that dictate the number of concentric layers \cite{Comolli2008,Chang2006,Effantin2006,Jiang2006} giving,  in turn, the effective diameter $d$ of the DNA. 
	We take the customary point of view that 
	the DNA molecule is a semiflexible elastic polymer characterized by the persistence length $L_p$ which is about the size of the radius of the capsid.  In the calculation of the bending modulus $k_3$,  we also assume that the hydrated DNA fills the entire volume of the capsid \cite{kellenberger1986considerations}, justifying our selection of the maximum linear density $m_0$.
	
	The repulsive energy between different DNA strands is not explicitly present in the model but is accounted for by the effective role of the parameters of the mechanical energy. 
	We construct the latter in two different stages. 
	We first consider the
	total energy of encapsidation consisting of the bending energy penalty of the DNA filament formulated in terms of the director field $\bn$,  together with a normalized potential $w(s)$  that attains its minimum value at the ordered configuration, labeled as $s=s_a\approx 1$,   and the maximum at $s=0$,   reflecting the preference of the DNA to organize inside the capsid.   A penalty is also associated with changes in $s$ resulting in a transition region between the coiled state and the melted core. 
	However, under  parameter choices that provide good agreement between the values of experimental and  the calculated disordered core sizes, we find that, the pressure is underpredicted by a factor of 100 with respect to experimentally measured values Finally,  pressure measurements for different viruses are also available \cite{cordova2003osmotic, Evilevitch2003, Grayson2006,   leforestier2009structure, Tzlil2003}. This leads us to 
	include an additional term in the energy penalizing distortions of the cross section formulated as quadratic expressions of the gradients of the  cross-sectional vectors $\nabla\bbm$ and $\nabla\bp$. The choice follows
from the	elastic energy
	of hexagonal columnar phases proposed by de Gennes. It is the bulk and shear modulus, B and C, present in the transverse elastic
	energy, that effectively account for the cohesive effects postulated in earlier models of DNA encapsidation \cite{Klug_JMPS2003},
\cite{Tzlil2003} that, in turn, effectively account for the electrostatic interactions.

	We find that the dimensionless energy restricted to spooling configurations  involves three material constants, $ \hat b, \hat \gamma_{0}$ and $\alpha$, in addition to $k_3$ that can be directly calculated from DNA data (Table 1). We determine the orders of magnitude of  $\hat b$ and $\hat\gamma_0$ from the  size of the disordered core of the T4 virus, as measured by imaging. 
 	 Likewise, the  parameter $\alpha, $  used in scaling the elastic moduli $B$ and $C$ with respect to $k_3$, is  estimated from experimental pressure values  corresponding to the same virus. The  numerical simulations   deliver phase spaces for the radius of the inner disordered core and the pressure, in terms of the parameters $ b$ and $\gamma_0$ (Figures 2-4).

	
	Earlier mathematical results on Ericksen's model guarantee existence of a global minimizer of the total energy \cite{Ambrosio_MM1990a}. In addition, numerical methods previously developed by Walker, et. al. \cite{nochetto2017finite,Nochetto_JCP2018,Walker_NM2018} on such a model extend to the present case and provide a convenient algorithm to predict packing configurations and the pressure values that they sustain. 
	\begin{figure}
		\centering{
			\includegraphics[width=3.1in]{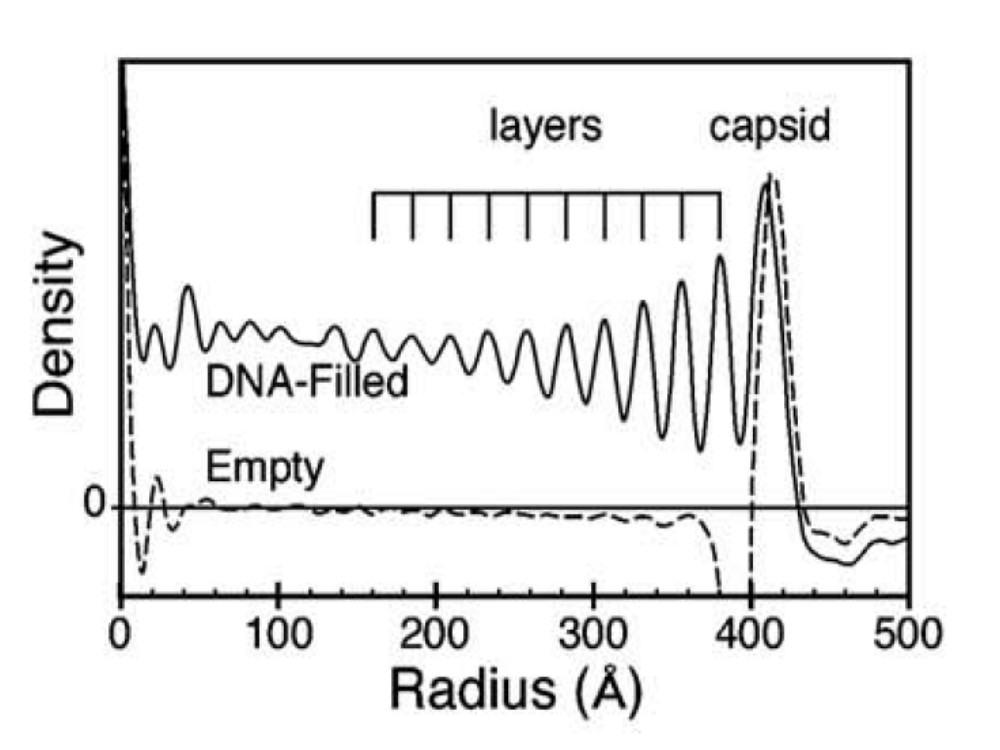}
			\includegraphics[width=3.3in]{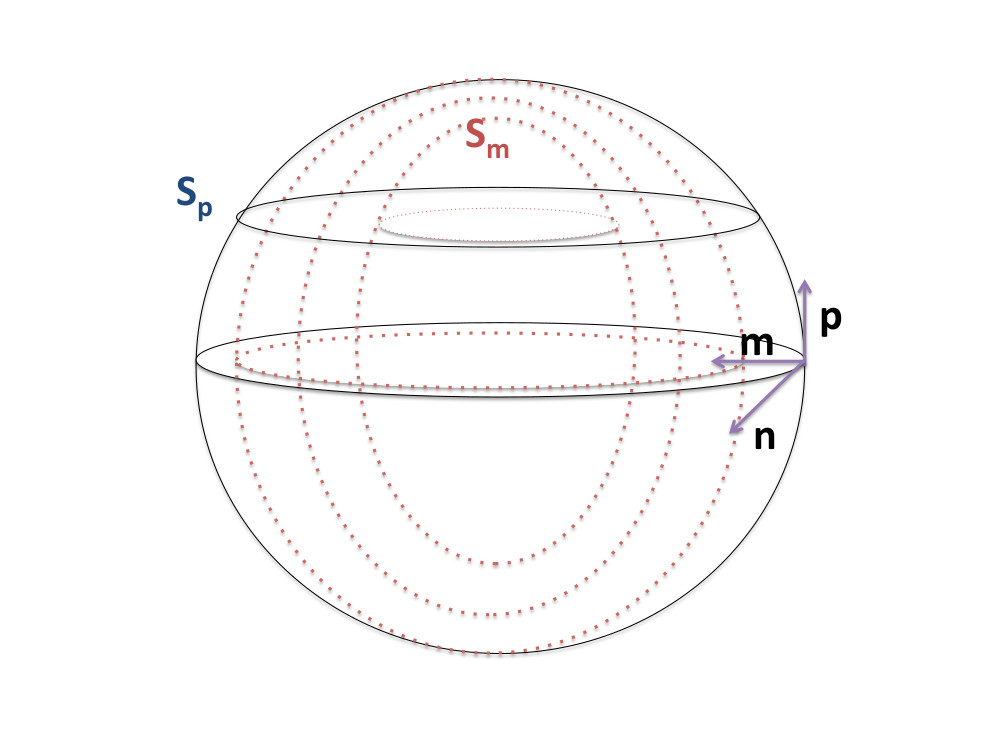}}
		\caption{The graph on the left represents the DNA density of bacteriophage $T5$. It shows the radial density distribution of DNA inside the viral capsid. The dashed line shows the density of an empty capsid. Figure reproduced from \cite{Molineux2013}. The figure on the right is a schematic representation of the level surfaces $\omega$=constant, \,$\vartheta$= constant, whose intersections provide the scaffolding that supports the DNA filament.}\label{fig:layer_exp_data}
	\end{figure}
Sine cryoelectron microscopy (cryo-EM) images  for most bacteriophages show multilayered spooling-like configurations on the outer layers of the packed genome \cite{Earnshaw1980,Cerritelli1997,Comolli2008,Chang2006,Effantin2006,leforestier2009structure,Jiang2006}, we present numerical simulations of minimizing configurations with spooling geometry.

	\begin{table}
		\begin{center}
			\begin{tabular}{|c|c|c|c|c|c|c|c|}
				\hline
				Virus  & $L_p$ (nm) & $d$ nm & $ L$ (nm)& $R_c$ (nm)& $r_0/R_c$ &Pressure(atm) \\ \hline\hline
				T4 & 52.8 & 2.4 &  55047.6 & 40.0 & 0.5500 & 28.70\\ \hline
				T5 & 58.38 & 2.94 & 39423.8 & 42.0 & 0.4286 & 30.17 \\ \hline
				T7 & 52.88 & 2.6 & 12932.0 & 26.05 & 0.5889 & 44.02 \\ \hline
				$\epsilon$15  & 53.9 & 2.55 &  12846.0 & 28.37 & 0.5735 & 40.41 \\ \hline
			\end{tabular}
			\caption{Physical measurements of four different bacteriophages.The symbol $L_p$ denotes the persistence length of a DNA chain of length $L$, effective diameter $d$, in a sphere-like capsid of radius $R_c$ with a measured radius $r_0$ of the disordered core. T4 \cite{leiman2003structure,olson2001structure}; T5 \cite{Effantin2006}; T7 \cite{Cerritelli1997}; $\epsilon$15 \cite{Jiang2006}; P22 \cite{Lander2006,Jin2015}.
			}
			\label{DNA-data-table}
		\end{center}
	\end{table}

	In section A, we  give a brief review of the nematic and chromonic models of liquid crystals, to explain and motivate our methods, with a special emphasis on Ericksen's model that includes the variable degree of orientation. 
	
	 In section B, we present the energy of packing, its scaling properties and the role of the spooling configurations. In section C, we present our approach for estimating  parameters, concluding with numerical simulations that relate  remaining   phenomenological parameters to the radius of the disordered core. 
	 Section D uses shape optimization tools to calculate the pressures near the capsid surface.  
	
This work is motivated by  earlier studies of free boundary problems of chromonic liquid crystals and bacteriophages carried out in the Ph.D dissertation by Lindsey Hiltner \cite{Lindsey2018}.

\subsection{Nematic liquid crystals}
In the Oseen-Frank theory, equilibrium states of a nematic liquid crystal are minimizers of the Oseen-Frank energy given by 
\begin{align}
E=&\int_{\Omega} \cW_1(\bn,\nabla\bn)\,d\bx\\
\cW_1 (\bn, \nabla \bn) =& k_1 (\Div \bn)^2 + k_2 (\bn \cdot \Curl \bn)^2 + k_3 |\bn \times \Curl \bn|^2 \nonumber\\
&\qquad + (k_2 + k_4) [\tr ( [\nabla \bn]^2 ) - (\Div \bn)^2]\label{Oseen-Frank}
\end{align}
The splay, twist and bend material constants, $k_1>0$, $k_2>0$ and $k_3>0$, respectively,  are  associated with the liquid crystal deformations that embody the corresponding geometries. The splay-bend term multiplying $k_2+k_4$ is a null Lagrangian and so, it depends on boundary values only. 
Defects in the Oseen-Frank theory correspond to locations, points, lines or surfaces, where $\bn=0$, that is where the molecular order is lost, giving rise to locally disordered structures, that may have infinite energy.

In order to represent liquid crystal configurations where ordered and disordered 
states coexist as well as assign finite energy to defects, it is convenient to introduce the order parameter $s$ in the model, and the energy associated with changes of order. For this, we will adopt Ericksen's model (1991)  of liquid crystals with variable degree of orientation. It may be viewed as an extension of the Oseen-Frank model; alternatively, it may also be interpreted as a restriction of that of Landau-deGennes to uniaxial states. 
In this model, the equilibrium state of the liquid crystal material is described by
the pair $(s, \bn)$ minimizing a bulk-energy functional
\begin{equation}\label{eqn:Erk_energy_general}
\begin{split}
\Eerk [s,\bn] &:= \frac{1}{2} \iO s^2 \cW_1 (\bn, \nabla \bn) + b|\nabla s|^2\,d\bx+ \Bulkcoef \iO \Bulkfunc (s)\,d\bx, \\
\end{split}
\end{equation} 
 with $\cW_1$ as in (\ref{Oseen-Frank}).
 The material constants $b>0$ and $\gamma_0>0$  are of the same of the general one, that contains coupling terms of 
 $\nabla\bn$ and $\nabla s$.
 (A full description of the model can be found in \cite{longa1987extension} and its analysis can be found in \cite{Nochetto_JCP2018, nochetto2017finite,Walker_NM2018}).
 
The double well function $\Bulkfunc(s)$ is of class $C^2$, defined on $-1/2 < s < 1$, that penalizes the phase transition between the nematic and the isotropic phases. It is chosen so that it has a minimum at 
  $s=s_a$, $s_a$ being the preferred order parameter value at a given temperature, or concentration; it has a maximum at $s=0$ indicating the ordering preference of the DNA.  We list its properties 
  as follows \cite{ericksen1991degree, Ambrosio_MM1990a, FLin_1991a}:
 \begin{enumerate}
 	\item $\lim_{s \rightarrow 1} \Bulkfunc (s) = \lim_{s \rightarrow -1/2} \Bulkfunc (s) = \infty$,
 	\item $\Bulkfunc(0) > \Bulkfunc(\Sas) = \min_{s \in [-1/2, 1]} \Bulkfunc(s) = 0$ for some $\Sas \in (0,1)$,
 	\item $\Bulkfunc'(0) = 0$.
 \end{enumerate}

Next, we discuss the approach to imposing boundary conditions. Rather than prescribing pointwise values of the fields $(\bn, s)$ on the boundary, that is, imposing strong anchoring boundary conditions,  we appeal to weak anchoring, which consists in penalizing the energy of departure of the boundary values from the ideal ones \cite{EVirga_1997}:
\begin{equation}\label{eqn:weak_anch_energy}
\begin{split}
\Eanch (s,\bn)& := \Ewan (s,\bn) + \Ewas (s), \\
\Ewan (s,\bn) &:= \frac{1}{2} \iG s^2 \left[ \aperp (\vnu \cdot \bn)^2 + \apar [1 - (\vnu \cdot \bn)^2] \right] dS(\bx),\\
\Ewas (s)& := \frac{1}{2} \iG \aori (s - \Sas)^2 dS(\bx),
\end{split}
\end{equation}
where $\vnu$ is the oriented unit normal vector of $\Gm := \partial \Om$. The first  term $\aperp$ tends to align the minimizing director field $\bn$ perpendicular  to $\vnu$, and the second term $\apar$ favors a parallel alignment. The last contribution to the energy 
is needed to ensure that $s$ does not trivially vanish on the interface, and so cause $\Ewan (s,\bn)$ to vanish as well \cite{Nochetto_JCP2018,Diegel_CCP2017,MorvantSealWalker_CAMWA2018}.
The parameters $\aperp, \apar, \aori : \Gm \to [0, \infty)$ may vary (in general), but here we take them to be fixed constants.

In determining a configuration of a liquid crystal, one minimizes the total energy consisting of (\ref{eqn:Erk_energy_general}) added to (\ref{eqn:weak_anch_energy}). A thorough study of existence of minimizer together with the numerical implementation of the minimizing algorithm can be found in \cite{Nochetto_JCP2018, nochetto2017finite, Walker_NM2018}. In particular, the theory allows for $\Omega$ to be of icosahedral shape, since $s=0$ at the edges where the vector fields are not defined.

\subsection{Chromonic Liquid Crystals}  We are concerned with hexagonal columnar phases of  chromonic 
liquid crystals. The energy consists of the Oseen-Frank contribution (or, the Ericksen's model analog)
together with an elastic energy of the cross-section perpendicular to the columnar director field $\bn$. The latter also effectively includes  the electrostatic repulsive energy of the negatively charged DNA strands. We then represent the energy as 
\begin{equation} \label{eqn:eneregy_chromonic}
\begin{split}
\Echr= \Eerk[\bn,s] + \Eelas[\nabla \bu],
\end{split}
\end{equation}
  where $\bu$ represents the displacement vector associated with deformation of the cross section. 
  The minimization of the chromonic energy is done subject to the constraints
  \begin{equation}\label{constraints}
  \nabla\cdot\bn=0, \quad \text{and}\quad (\nabla\times\bn)\cdot\bn=0.
  \end{equation}
The splay of a liquid crystal is zero whenever dislocations do not occur, that is, the same number of filaments that enter a unit area exit that cross-section. 
In the case of the hexagonal columnar phase, nonzero splay would allow for deviations from the lattice structure.
Twist is prohibited because of its incompatibility with the two-dimensional lattice order in planes perpendicular to the director.

We conclude this section with the observation that the previously models described  correspond to a continuum and do not reflect the discrete nature of the system  made of filaments. These do not overlap except, perhaps, in possible knot locations. In order to associate a minimizing set $\{\bn, \mm, \bp, s\}$ to a DNA configuration, given the inter-strang DNA spacing taken as its effective diameter, $d>0$,  we further introduce the scalar fields $\omega, \vartheta$ as solutions of the differential equations
\begin{equation}\label{eqn:level_curves}
 \frac{d\omega}{d\boldsymbol m}=\frac{2\pi}{d}, 
 \quad \frac{d\vartheta}{d\bp}=\frac{2\pi}{d}, \quad s\neq 0.
\end{equation}
Then, the intersections of level surfaces $\omega=\text{constant}$, $\vartheta=\text{constant}$ give the location of the DNA strands. Furthermore, the piecewise smooth curve  $\br=\br(\zeta), \zeta\in [0,L]$,
such that \begin{equation} \label{eqn:central axis}
\frac{d\br}{d\zeta}=\bn(\br(\zeta)), \quad \br(0)=\br_0,
\end{equation}
  describes the axis of the DNA. Here $L>0$ represents the length of the genome and $\br_0$ its entry point in the capsid. Restricting the elastic energy in (\ref{eqn:eneregy_chromonic}) to discrete deformations of the level curves sustaining the DNA filaments, we arrive at the expression
  \begin{equation}
  \Eelas= B|\nabla(\mm+\bp)|^2 + C|\nabla(\mm-\bp)|^2,
  \end{equation} 
where $B$ and $C$ correspond to bulk and shear moduli, respectively. 
\section{DNA encapsidation}\label{sec:DNA_encapsidation}
%
We assume that the capsid corresponds to a bounded, axisymmetric region
$\Omega\in \mathbb R^3$ with  $\partial \Omega$
representing the piecewise smooth, faceted, viral capsid. We let $l_0>0$ denote the length of the axis that connects the location of the connector with its antipodal site in the viral capsid. Assuming the DNA is tightly packed inside the capsid, we adopt a continuum approach for representing the configuration of the DNA \cite{Klug_JMPS2003,Klug_CM2005}.  To this end, we let $\bn$ represent the unit tangent vector of an ``ensemble'' of packed DNA strands.  Moreover, we use the $s$ variable (degree-of-orientation) to represent how well ordered the DNA strands are, e.g. $s=0$ indicates that the DNA strands are oriented in all directions equally. Furthermore, the unit vectors $\{\mm, \bp\}$ span the cross sections perpendicular to the DNA axis, with its associated level curves representing strand 
locations. 

\subsection{Total energy}
The model that we propose consists of the total free energy together with a list of constraints, expressed as follows:
\begin{align}
E =&\Eerk [s,\bn]+ \Eanch (s,\bn)  + \int_{\Omega}[s^2\big(B|\nabla(\bbm+\bp)|^2 + C|\nabla(\bbm-\bp)|^2 \big)] \, d\bx, \label{Ecapsid}\\
&\bn\cdot\bbm=0, \, \, \bbm\cdot\bp=0, \, \, \bn\cdot\bp=0. \quad |\bn|=|\bbm|=|\bp|=1, \label{orthonormal-vectors}  
\end{align}

 We make the following choices for the parameters of the model:
\begin{align}
k_1 &\approx k_2 \gg k_3, \quad k_4 = 0, \label{eqn:ki} \\
b &\approx k_3, \quad B\approx k_3 \approx C, \quad A:=B+C=\alpha k_3,\label{eqn:bBC}\\
\anchcoef& := \aperp = \aori, \quad \apar = 0, \quad \anchcoef \approx k_3 / R_c, \label{eqn:a}
 \end{align}
where $\alpha>0$ in (\ref{eqn:bBC}) is a scaling parameter latter determined using pressure measurements. 
The choices for $\{ k_i \}_{i=1}^4$ indicate that $\Div \bn$ and $\bn \cdot \Curl \bn$ are almost negligible for the DNA coiling configurations that we consider here. Dimensional analysis as well as information derived from Onsager's theory of rigid rods will further assist us in specifying  relations (\ref{eqn:bBC}). The selection of the anchoring constants (\ref{eqn:a}) enforces tangential (planar)  boundary conditions for $\bn$, as this is consistent with DNA coiling inside capsids.  

The bulk potential function $\Bulkfunc(s) : [0,1] \to [0,\infty)$ is modified to satisfy:
\begin{equation}\label{eqn:bulk_pot_DNA}
\begin{split}
\Bulkfunc(\Sas) &= \min_{0 \leq s \leq 1} \Bulkfunc(s) = 0, \qquad 	\Bulkfunc \text{ is quadratic and convex near } s = \Sas, \\
\Bulkfunc(0) &= 1, \qquad \Bulkfunc \text{ has a local max at } s = 0,
\end{split}
\end{equation}
The optimal degree of orientation is chosen to be $\Sas := 0.8$.


\color{black}

%

The total energy of the DNA coiling system is
\begin{equation}\label{eqn:DNA_coiling_energy}
\begin{split}
\Energy [s,\bn,\bbm,\bp,\Om]
&= \frac{1}{2} \iO s^2 \big{[} k_1 (\Div \bn)^2 + k_2 (\bn \cdot \Curl \bn)^2 + k_3 |\bn \times \Curl \bn|^2 \\
&\qquad\quad + k_2 [\tr ( [\nabla \bn]^2 ) - (\Div \bn)^2] \big{]} d\bx + \frac{1}{2} \iO b |\nabla s|^2 d\bx \\
&\quad + \Bulkcoef \iO \Bulkfunc (s) \, d\bx + \frac{1}{2} \anchcoef \left[ \iG s^2 (\vnu \cdot \bn)^2 dS(\bx) +  \iG (s - \Sas)^2 dS(\bx) \right] \\
&\quad + \frac{1}{2} \iO \left[B s^2 |\nabla (\bbm + \bp)|^2 + C s^2 |\nabla (\bbm - \bp)|^2 \right] d\bx.
\end{split}
\end{equation}

\subsection{Model simplification: spooling configurations}\label{sec:simplify}
From now on, we consider axisymmetric capsids and spooling configurations. Adopting cylindrical coordinates $(r, \theta, z)$,  with the $z$-axis along the capsid, spooling DNA arrangements are characterized by $\bp = \bbe_z$. Furthermore, for such configurations,  it is natural to set $\mathbf \nu= \be_r$ and $\bn = \be_{\theta} $ and $s=s_a$ on the boundary. 
 Since $\nabla\bp=0$ and $\bbm = \bn \times \bp$, so  $|\nabla \bbm| = |\nabla \bn|$.  
 We now define ${\tilde k}_i = k_i + A, \,\, i=1, 2, 3,$ with $A$ as in (\ref{eqn:bBC}). In particular,
 note that 
 \begin{equation}
 {\tilde k}_3= k_3(1+\alpha). \label{tilde-k3}
 \end{equation}
  Finally, using the identity:
\begin{equation*}
(\bn \cdot \Curl \bn)^2 + |\bn \times \Curl \bn|^2 + \tr ( [\nabla \bn]^2 ) = |\nabla \bn|^2,
\end{equation*}
	   we obtain
\begin{equation}\label{eqn:DNA_coiling_energy_simplify_1}
\begin{split}
\Energy [s,\bn,\Om] &= \frac{\tilde k_3}{2} \iO s^2 \widetilde{\cW}_1 (\bn, \nabla \bn) d\bx + \frac{1}{2} \iO b |\nabla s|^2 d\bx + \Bulkcoef \iO \Bulkfunc (s) \, d\bx \\
 +&\frac{1}{2} \anchcoef \left[ \iG s^2 (\vnu \cdot \bn)^2 dS(\bx) +  \iG (s - \Sas)^2 dS(\bx) \right]
\end{split}
\end{equation}
with
\begin{equation}
\begin{split}
\widetilde{\cW}_1 (\bn, \nabla \bn) &= (\tilde k_1/\tilde k_3) (\Div \bn)^2 + (\tilde k_2/\tilde k_3) (\bn \cdot \Curl \bn)^2 + |\bn \times \Curl \bn|^2 \\
&\qquad + (\tilde k_2/\tilde k_3) [\tr ( [\nabla \bn]^2 ) - (\Div \bn)^2].
\end{split}
\end{equation}

\subsection{Dimensional Analysis}\label{sec:non_dim}

Introducing a length scale $L_0$ to be of the same order as $R_c$, and noting the following scalings:
\begin{equation*}
\Om = L_0^3 \widehat{\Om}, \quad \Gamma = L_0^2 \widehat{\Gamma}, \quad \partial \Om_0 = L_0^2 \widehat{\partial \Om_0}, \quad \nabla = \frac{1}{L_0} \widehat{\nabla},
\end{equation*}
where $~\widehat{ }~$ denotes non-dimensional quantities, \eqref{eqn:DNA_coiling_energy_simplify_1} becomes
\begin{equation}\label{eqn:DNA_coiling_energy_non_dim}
\begin{split}
\Energy [s,\bn] &= \tilde{k}_3 L_0 \widehat{\Energy} [s,\bn] \\
\widehat{\Energy} [s,\bn] &:= \frac{1}{2} \iO s^2 \widetilde{\cW}_1 (\bn, \nabla \bn) \, d\bx + \frac{\hat{b}}{2} \iO |\nabla s|^2 d\bx + \Bulkhatcoef \iO \Bulkfunc (s) d\bx \\
&\qquad\qquad\quad + \frac{\anchhatcoef}{2} \left[ \iG s^2 (\vnu \cdot \bn)^2 dS(\bx) + \iG (s - \Sas)^2 dS(\bx) \right], \\
\end{split}
\end{equation}
where
\begin{equation}\label{eqn:DNA_coiling_energy_non_dim_constants}
\begin{split}
\hat{b} = \frac{b}{\tilde{k}_3}, \quad \Bulkhatcoef = \frac{L_0^2}{\tilde{k}_3} \Bulkcoef, \quad \hat k_i=\frac{\tilde k_i}{\tilde k_3}, \, i=1,2, \quad  \anchhatcoef = \frac{\anchcoef L_0}{\tilde{k}_3} = \frac{100 L_0}{R_c},
\end{split}
\end{equation}
and we have dropped $~\widehat{ }~$ on all variables for convenience (the domains are now non-dimensional). 

\section{Parameter estimates and numerical simulations}\label{sec:est_constants}
As for semiflexible polymers and its application to the DNA molecule, following Tzlil et al. \cite{Tzlil2003} and \cite{Klug_JMPS2003}, we take the bending constant $k_3$ as
\begin{equation}
k_3= K_B T L_p m_0. \label{K3}
\end{equation}
The DNA cross-sectional length density, $m_0$, is estimated by taking the ratio of the length of a DNA strand to the volume of the container in which it is packed.  Thus, for a perfectly packed DNA strand (i.e. no empty space in the volume), the \emph{maximum value} is $m_0 = 1 / (\pi (d/2)^2) = 1.85131 \times 10^{17}$ m$^{-2}$, where we used an average DNA diameter of $d = 2.6225$ nanometers from Table \ref{DNA-data-table}.
More realistically, for an average bacteriophage genome (length $30.0624$ micrometers) and average capsid head (radius $34.11$ nanometers), we get an average $m_0 = 1.809173 \times 10^{17}$ m$^{-2}$.  Furthermore, taking an average persistence length of $L_p = 54.49$ nanometers, our effective bending constant is $k_3 = 4.058027 \times 10^{-11}$ J/m (assuming room temperature).

It is convenient to distinguish three types of parameters in the energy. First of all, $\tilde k_3$ appears as a multiplicative constant whose role will be adjusting the pressure.
The ratios $\tilde k_1/\tilde k_3, \tilde k_2/\tilde k_3$ should be interpreted as having large values to penalize the constraints (\ref{constraints}). The ratio $\hat{a}_0 = a_0 L_0 / \tilde k_3$ is used to enforce the planar boundary conditions, so it also has a large penalty value.

\begin{figure}
	\centerline{
		\includegraphics[width=3.3in]{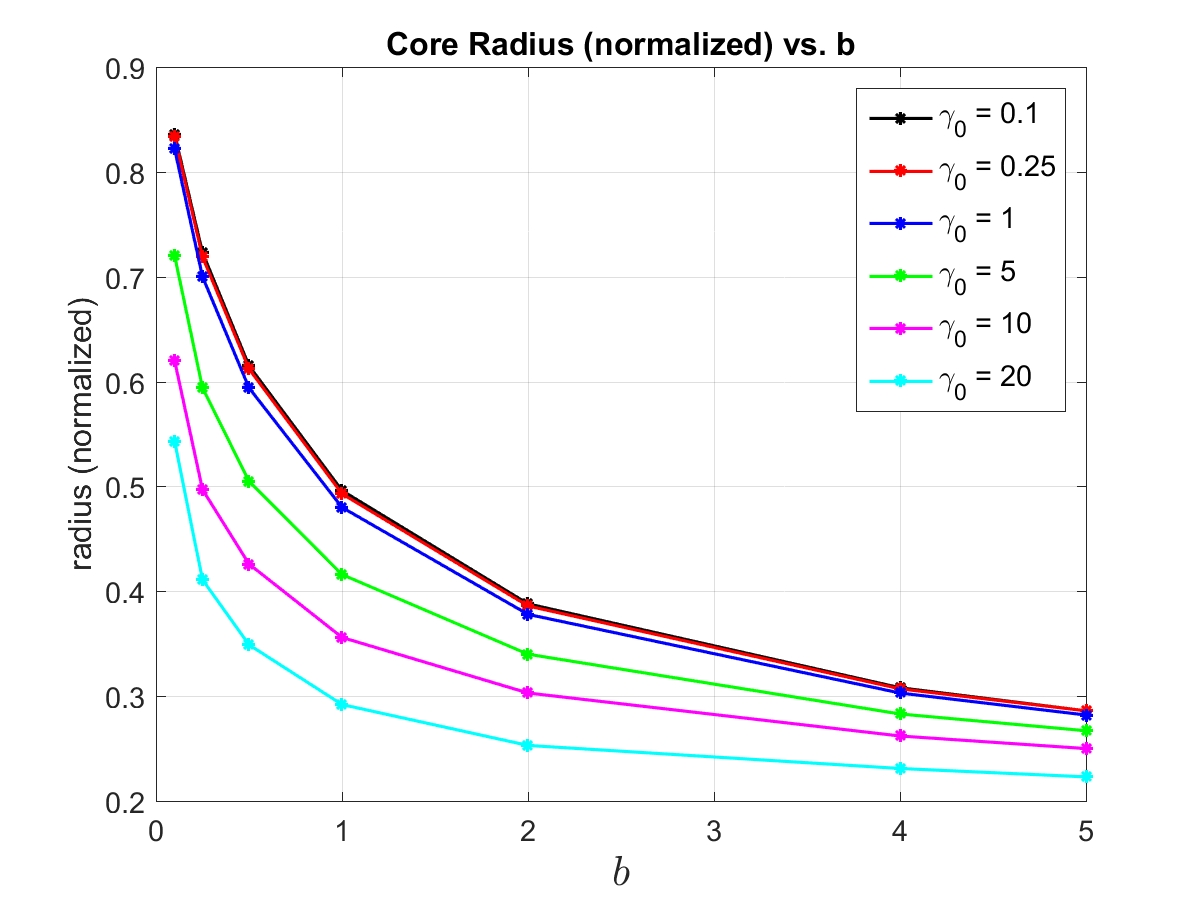}\quad \quad  \quad 
		\includegraphics[width=3.2in]{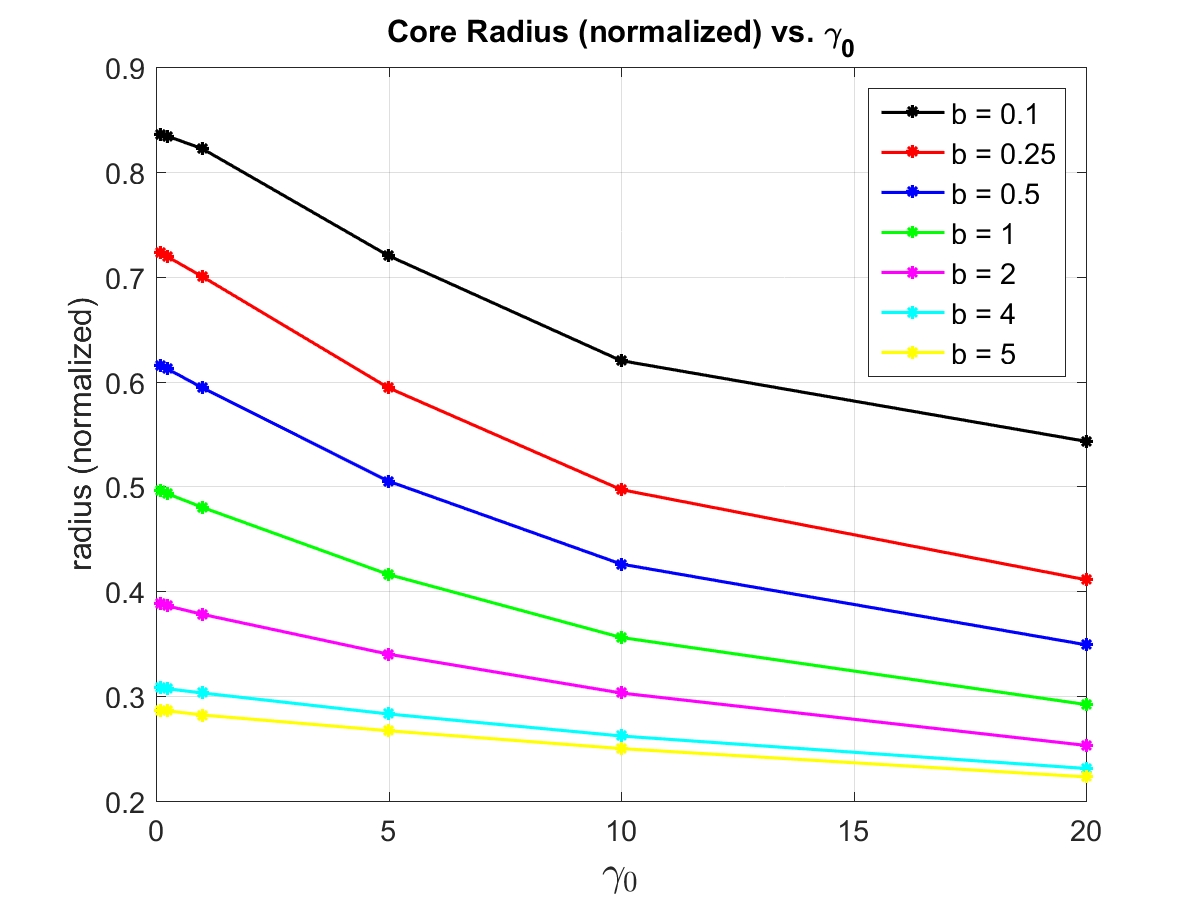}}
	\caption{The graphs show the role of the parameters $b$ and $\gamma_0$ in the model. In both cases, the  curves indicate the decrease in core radius with increasing parameter values,  as a mechanism to lower the energy of the system. }
	\label{core_radius_vs_b_1_gamma_0}
\end{figure}

\begin{figure}
	\centerline{
		\includegraphics[width=3.3in]{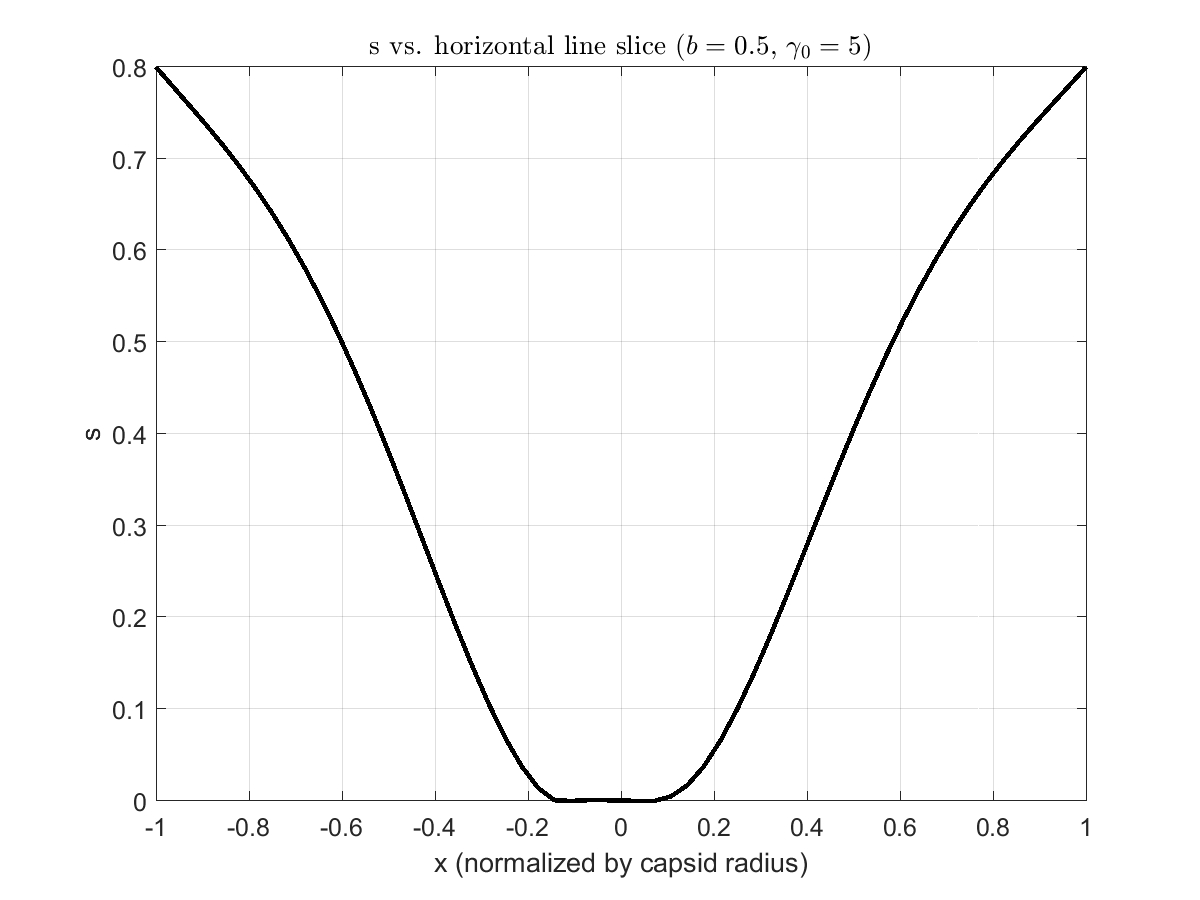}\quad \quad  \quad 
		\includegraphics[width=2.2in]{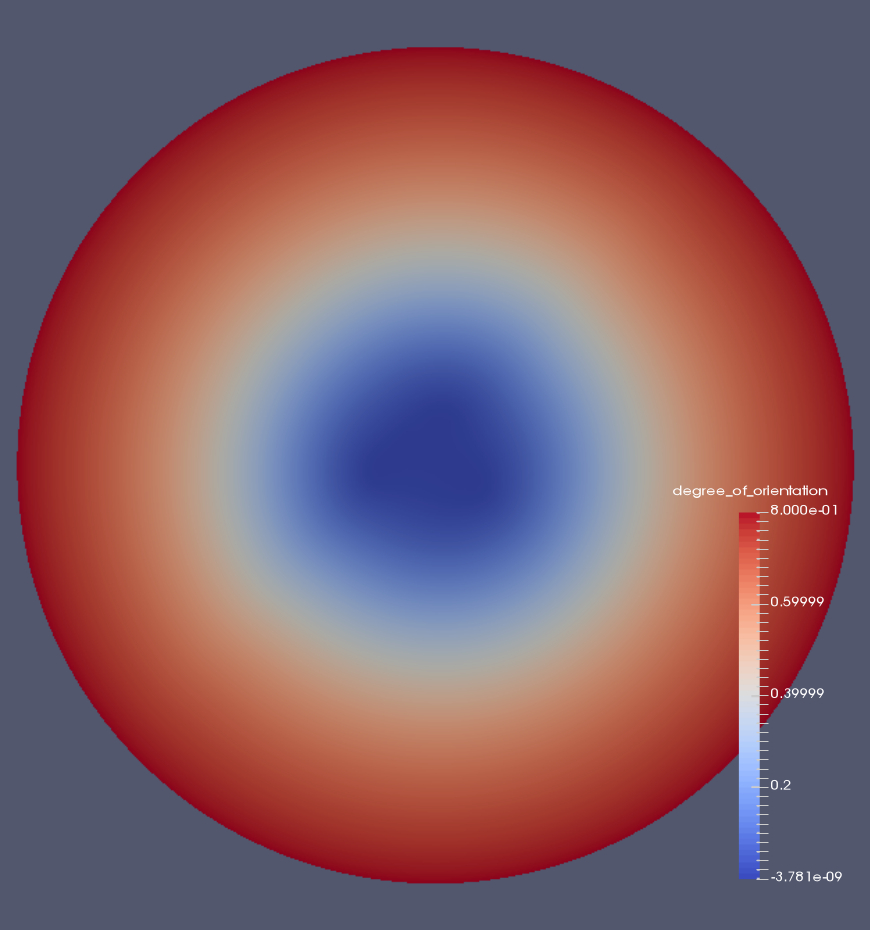}}
	\caption{The graphs show two different profiles of the order parameter in the capsid domain. 
	Left plot is of $s$ evaluated along a line through the equator of the capsid.  The center of the capsid is located at 0, and the capsid radius is normalized to 1. Since the maximum order corresponds to $s=0.8$, we take the disordered threshold at $s=0.4$, which gives the ratio $r_0/R_c \approx 0.496$. Right plot: shows the order profile of the capsid transitioning from order DNA (red) to disordered (blue) as a 2-D slice through the capsid equator.}
	\label{fig:horizontal_line_slice}
\end{figure}

For computations, we take $\tilde{k}_1 = \tilde{k}_2 = 10 \tilde{k}_3$, where $\tilde{k}_3 = k_3 (1 + \alpha)$ (recall \ref{eqn:bBC}) and $\alpha > 0$ is determined by the pressure measurements (see Section \ref{sec:shape_optim}). The length scale $L_0 = R_c / 0.4$, where $0.4$ is the non-dimensional radius of the computational capsid domain; thus, $\anchhatcoef = 250$.
Moreover we explore the parameter range for $\hat{b}$ and $\Bulkhatcoef$ given by:
\begin{equation}\label{eqn:non_dim_const_range}
\begin{split}
	0.1 \leq \hat{b} \leq 5.0, \qquad  0.1 \leq \Bulkhatcoef \leq 20,
\end{split}
\end{equation}
and compare the predicted cores sizes with the available virus measurements shown in Table \ref{DNA-data-table}. The outcome from the numerical simulations is summarized in Figure \ref{core_radius_vs_b_1_gamma_0}.  Figure \ref{fig:horizontal_line_slice} shows how the internal ``core radius'' was estimated for a particular choice for $b$ and $\gamma_{0}$.  Figure \ref{pressure_graphs} shows how the capsid pressure, and the core radius, varies with $b$ and $\gamma_{0}$.

\section{Shape optimization}\label{sec:shape_optim}

The minimum energy state of the packing problem is the one that minimizes $\Energy$ in \eqref{eqn:DNA_coiling_energy_non_dim} with respect to $s$ and $\bn$.  With this, one can compute pressures in the capsid by perturbing $\Energy$ with respect to the domain $\Om$, i.e. take shape derivatives \cite{Walker_book2015}.

\begin{figure}
	\centerline{
		\includegraphics[width=3.3in]{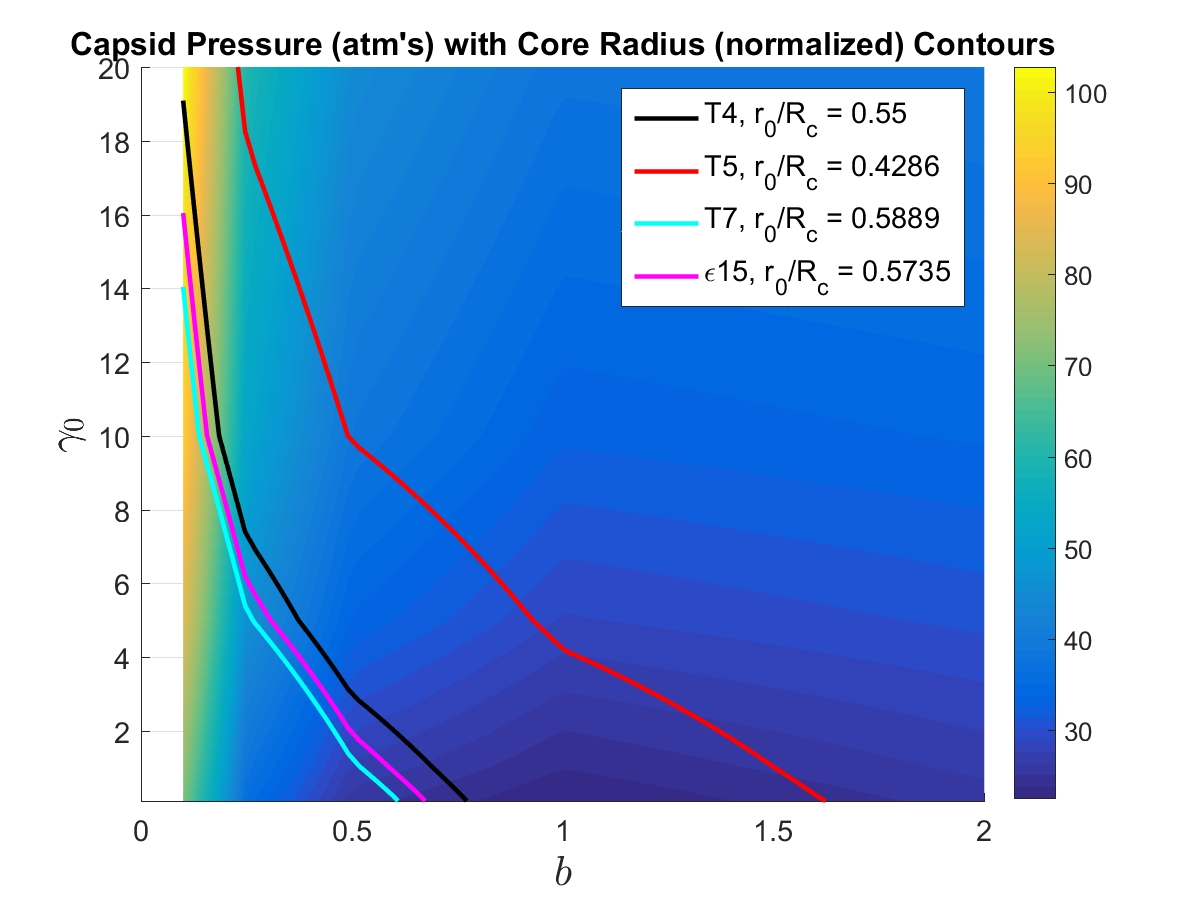}\quad \quad  \quad 
		\includegraphics[width= 3.7in]{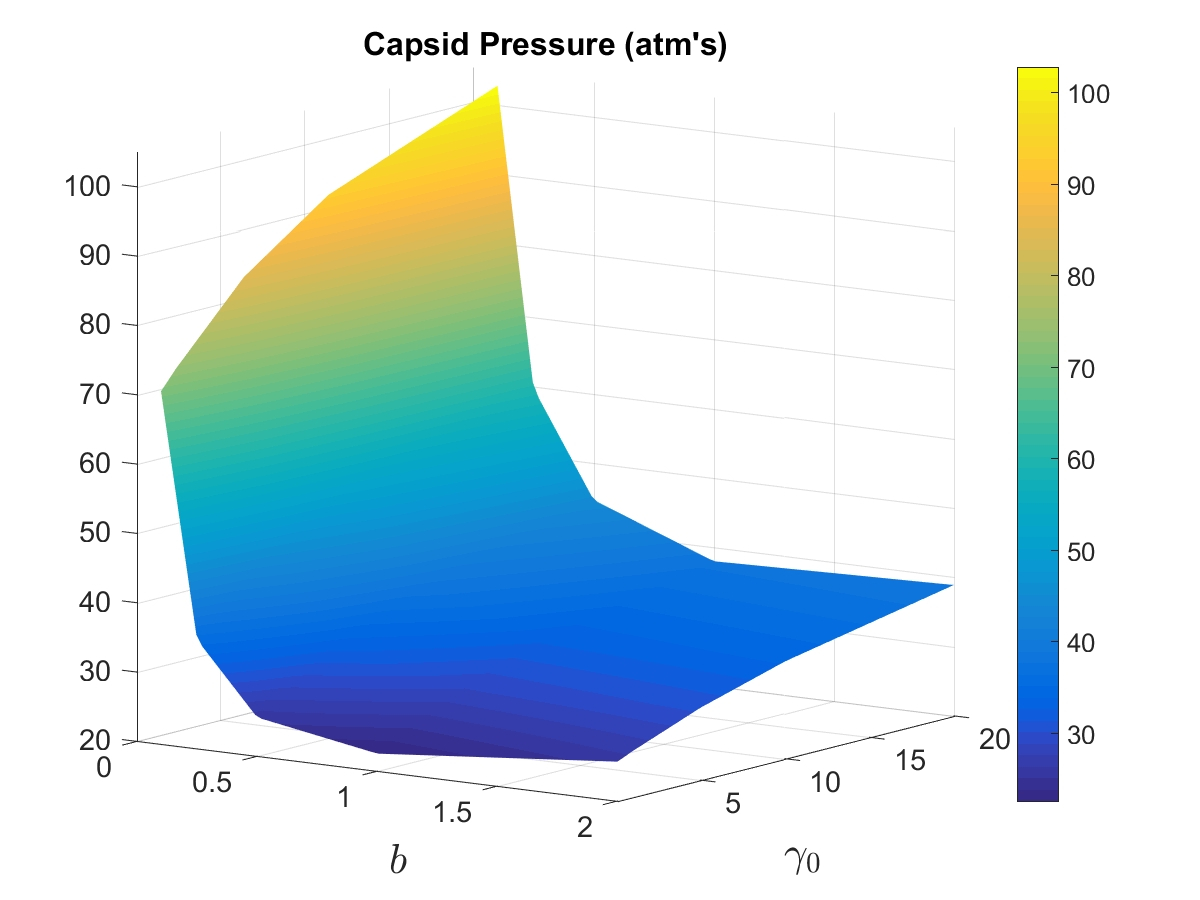}}
	\caption{Plots of core radius and capsid pressure.  Left plot show curves of constant core radius for different choices of $b$ and $\gamma_0$ overlaid on the pressure values.  In the region where the core radius matches experiment, the pressure values are also comparable to experimental values (from a minimum of about $26$ atms to a maximum of about $95$ atms).  The right plot is a 3-D view of the pressure. Note: $\alpha$ is chosen as $75$, which fully determines $\tilde{k}_3$, to give a decent fit to all four viruses.}
	\label{pressure_graphs}
\end{figure}

More specifically, given a (non-dimensional) perturbation field $\vV$ that deforms $\Om$, the shape variation $\delta_{\Om} \widehat{\Energy} [s,\bn,\Om; \vV]$ gives the (non-dimensional) change in energy due to a change in the volume. 
For example, let $\vV_{\Gm}$ be a smooth vector field such that:
\begin{equation}\label{eqn:vV_outer_inner_bdy}
\begin{split}
	\vV_{\Gm} &= \bnu, \text{ on } \Gm, \text{ and smoothly decays to zero inside } \Om.
\end{split}
\end{equation}
Then, the dimensional change in energy due to expanding the outer boundary $\Gm$ is $\tilde{k}_3 \delta_{\Om} \widehat{\Energy} [s,\bn,\Om; \vV_{\Gm}]$ which has units of force: $J/m = N$.
Thus, the average (dimensional) pressure being exerted on the boundary of the capsid $\Gm$ is
\begin{equation}\label{eqn:ave_pressures}
\begin{split}
	\text{outer pressure } &= \frac{1}{L_0^2 |\Gm|} \tilde{k}_3 \delta_{\Om} \widehat{\Energy} [s,\bn,\Om; \vV_{\Gm}],
\end{split}
\end{equation}
where we note that $|\Gm|$ is the non-dimensional surface area of the capsid.  Note that $\vV_{\Gm}$ is normalized so that $|\vV_{\Gm}|=1$ on $\Gm$.  The particular form of the extension in \eqref{eqn:vV_outer_inner_bdy} does not matter because ultimately the shape derivative only depends on what happens at the boundary $\Gm$ by the Hadamard-Zol\'{e}sio structure theorem \cite{Walker_book2015}. The graphs of the pressure in terms of the parameters $b$ and $\gamma_0$ are shown in Figure \ref{pressure_graphs}.

\section{Conclusions} We constructed a model of encapsidation of viral DNA, that combines that of hexagonal chromonic liquid crystals with the theory of liquid crystals with variable degree of orientation. 
The model is accompanied by a rigorous, finite-element based, algorithm of energy minimization, with design and prediction capabilities that may prove to be a real asset in applications. Due to the presence of the order parameter, the algorithm is able to deal with very general faceted capsid domains, including the geometry of the packing motor features, and also allow for the presence of knots in the DNA.  The outcome of the model is a phase space for the radius of the disordered core and the pressure near the boundary of the capsid, with the potential to aid in the design of viral DNA to meet desired specifications.  The model also allows for  the exploration of 
 large sets of parameters, reiterating its value as a design tool.  Follow up work addresses the presence of ions and the dynamics of DNA ejection from the capsid. 
 

\bibliography{pnas}
\bibliographystyle{plain}

\end{document}